# Kondo physics in antiferromagnetic Weyl semimetal Mn$_{3+x}$Sn$_{1-x}$ films

Durga Khadka,[1,#] T. R. Thapaliya,[1,#] Sebastian Hurtado Parra,[2] Xingyue Han,[2] Jiajia Wen,[3] Ryan F. Need,[4,5] Pravin Khanal,[6] Weigang Wang,[6] Jiadong Zang,[7] James M. Kikkawa,[2] Liang Wu,[2] S. X. Huang[1,*]

[1]Department of Physics, University of Miami, Coral Gables, Florida, 33146, USA.
[2]Department of Physics and Astronomy, The University of Pennsylvania, Philadelphia, PA 19104, USA.
[3]Stanford Institute for Materials and Energy Sciences, SLAC National Accelerator Laboratory, Menlo Park, California 94025, USA.
[4]Department of Materials Science and Engineering, University of Florida, Gainesville, Florida, 32611, USA.
[5]NIST Center for Neutron Research, National Institute of Standards and Technology, Gaithersburg, Maryland 20878, USA.
[6]Department of Physics, University of Arizona, Tucson, AZ 85721, USA.
[7]Department of Physics, University of New Hampshire, NH, 03824, USA.

[#]These authors contributed equally to this work
[*]Corresponding author. Email: sxhuang@miami.edu

**Abstract:**

Topology and strong electron correlations are crucial ingredients in emerging quantum materials, yet their intersection in experimental systems has been relatively limited to date. Strongly correlated Weyl semimetals, particularly when magnetism is incorporated, offer a unique and fertile platform to explore emergent phenomena in novel topological matter and topological spintronics. The antiferromagnetic Weyl semimetal Mn$_3$Sn exhibits many exotic physical properties such as a large spontaneous Hall effect and has recently attracted intense interest. In this work, we report synthesis of epitaxial Mn$_{3+x}$Sn$_{1-x}$ films with greatly extended compositional range in comparison with that of bulk samples. As Sn atoms are replaced by magnetic Mn atoms, the Kondo effect, which is a celebrated example of strong correlations, emerges, develops coherence, and induces a hybridization energy gap. The magnetic doping and gap opening lead to rich extraordinary properties as exemplified by the prominent DC Hall effects and resonance-enhanced terahertz Faraday rotation.

# INTRODUCTION

Band structure topology is becoming an increasingly important aspect in materials research and design, and is responsible for many of the exotic behaviors found in quantum materials such as strongly-correlated 2D electron systems (*1*), graphene (*2*) and topological insulators (*3*). More recently, band structure topology has been critical to the understanding and development of gapless topological semimetals (TSMs) (*4-6*) including: Weyl semimetals (WSMs), Dirac semimetals, and nodal line semimetals. In Weyl semimetals (WSMs), the conduction and valence bands cross at certain points in momentum (*k*-) space known as Weyl nodes (*6*). Weyl nodes appear in nondegenerate pairs with opposite chirality and act as monopoles of Berry curvature (i.e. sources or sinks of Berry flux). The spacing of the Weyl nodes in *k*-space is determined by the crystal and/or magnetic structure that break inversion or time-reversal symmetry, and in turn dictates the magnitude of the intrinsic anomalous Hall effect (*7, 8*).

While much of the work on WSMs to date has focused on weakly interacting systems, there is a growing need to address and include the effects of strong electron correlations (*9-12*). A classic example of strongly correlated behavior is the Kondo effect (*13*), which originates from the coupling between the spins of conduction electrons and local magnetic moments. As local magnetic moments form a lattice, the Kondo effect develops coherence and leads to the Kondo insulator (*14, 15*). Inclusion of topological order in Kondo insulators leads to *topological* Kondo insulators, among which, $SmB_6$ is an example that has attracted significant attention (*16*). Recently, there have been increased theoretical efforts (*17-19*) to address the Kondo effect in WSMs and particularly the emergence of a Weyl-Kondo semimetal in non-magnetic $Ce_3Bi_4Pd_3$ bulk samples (*20*). Unusual properties such as giant spontaneous Hall effect were predicted (*21*). These works suggest that strongly correlated WSMs are a fertile platform to explore new quantum phases resulting from the interplay between Weyl and Kondo physics.

A promising material in which to study both Weyl and Kondo physics is the antiferromagnetic (AFM) WSM $Mn_3Sn$ (*22-25*). $Mn_3Sn$ has hexagonal structure with lattice constants $a = 5.67$ Å and $c = 4.53$ Å (Fig. 1A), with a Néel temperature ($T_N$) of 420 K (*25*). The Mn atoms form a 2D Kagome lattice on (0001) plane (i.e., *ab*-plane) with Sn atoms sitting at the hexagon centers (Fig. 1B). Two neighboring Kagome $Mn_3Sn$ layers are stacked along *c*-axis with an in-plane lattice offset (Fig. 1A). The Mn moments (~3 $\mu_B$/Mn) form various non-collinear spin orders at different temperatures including inverted triangular structures, spin spiral, and spin glass (*26*) (Fig. S1, Supplementary Note S1). The most interesting are the inverted triangular spin textures (Fig. 1B) with nearly zero spontaneous magnetization (~$3\times10^{-3}$ $\mu_B$/Mn). The magnetic symmetry of the inverted triangular spin textures includes mirror symmetry that ensures the appearance of the pairs of Weyl nodes at a *k* line along the local easy axis in the *ab*-plane, as demonstrated by the ARPES measurements (*22*). The Berry curvature induced by the Weyl nodes generates a large anisotropic anomalous Hall effect (AHE) (*22, 25*). Under an external field, tilting of the triangular spin textures results in non-coplanar spin structures and generates a topological Hall effect arising from the spin Berry curvature in real space (*27*). Moreover, novel magnetic spin Hall and inverse spin Hall effects, which demonstrate field-controllable conversion between spin current and charge current, have been discovered (*24*). In addition to these unusual topological and spintronic properties, significant bandwidth renormalization is observed due to strong correlations among Mn 3*d* electrons (*22*). These outstanding topological/spintronic properties and strong electron correlations make $Mn_3Sn$ an ideal platform to explore the multifaceted physics of the interplay between topology, magnetism, and strong correlations, as well as the emerging topological antiferromagnetic spintronics (*28, 29*) in which thin films play a crucial role.

Here we report the fabrication of high-quality epitaxial films of $Mn_{3+x}Sn_{1-x}$ with greatly extended compositional range ($x$ up to ~ 0.55). We observe the Kondo effect in films with excess Mn, which substitutes for Sn and acts as a magnetic dopant in the system. Upon increased Mn doping, the system develops Kondo coherence and opens a hybridization gap. Films with $x > 0.4$ behave like Kondo insulators with topological features, as manifested in the DC Hall and terahertz (THz) Faraday rotation measurements.

## RESULTS

### Fabrication of epitaxial $Mn_{3+x}Sn_{1-x}$ films with greatly extended compositional range

$Mn_3Sn$ has layered Kagome planes and exhibits a strongly anisotropic Hall effect. Thus, it is important to realize epitaxial films with two different orientations: (0001) films where the Kagome plane is in the plane of the film, and $(11\bar{2}0)$ films where the Kagome plane is perpendicular to the plane of the film. There are some reports on the fabrication of near-stoichiometric $Mn_3Sn$ films, including (0001) $Mn_3Sn$ films on Ru seed layer (*30*), (0001) and $(11\bar{2}0)$ $Mn_3Sn$ films on MgO and $Al_2O_3$ substrates (*31*), and polycrystalline films on $Si/SiO_2$ substrate (*32*). In this work, we used co-sputtering of Mn and Sn targets and realized epitaxial growth of (0001) and $(11\bar{2}0)$ $Mn_{3+x}Sn_{1-x}$ films (see Materials and Methods).

We took advantage of the small lattice mismatch (~2%) between Pt (111) (cubic, $\sqrt{2}a = 5.55$ Å) and (0001) $Mn_3Sn$ ($a = 5.67$ Å), and fabricated high quality epitaxial (0001) $Mn_{3+x}Sn_{1-x}$ films on a Pt seed layer, which is grown on $C-Al_2O_3$ substrate via magnetron sputtering. The $\theta/2\theta$ x-ray diffraction (XRD) pattern (Fig. 2A) shows a strong (0002) $Mn_3Sn$ peak without any noticeable impurity peaks. Importantly, high resolution XRD (HR-XRD) rocking curves, collected with monochromating incident optics and a triple axis analyzer, show a very sharp peak with full width half maximum (FWHM) of 0.002° (Fig. 2B, for $x = 0.5$), indicating the film has exquisite planar alignment between (0001) crystalline plane and film plane. Rocking curves with a double axis analyzer showed a sharp peak with a slightly larger FWHM of 0.07° (Fig. S2A), indicating a small $d$-spacing spread. For Sn-rich samples (e.g., $x = 0.1$), crystalline quality decreases slightly with a rocking curve FWHM of 0.21° (double axis, Fig. S2B) and in-plane $\varphi$ scan FWHM of 0.8° (Fig. S2C). In-plane $\varphi$ scans (Fig. 2C) of $Mn_3Sn$, Pt, and $Al_2O_3$ show 6 diffraction peaks confirming the epitaxial relation of $Al_2O_3$ $[10\bar{1}0]$ || Pt [110] || $Mn_3Sn$ $[2\bar{1}\bar{1}0]$. Moreover, atomic force microscopy measurements indicate the RMS value of the surface roughness (Fig. S2D) is only about 0.4 nm. The flat surface and sharp interface between Pt and (0001) $Mn_3Sn$ are further confirmed by x-ray reflectivity measurements (Fig. S2E). We attribute the high quality of $Mn_3Sn$ to the high quality Pt seed layer as indicated by the Laue oscillations of the XRD Pt (111) peak (inset of Fig. S2E). We note that the flat surface of (0001) $Mn_3Sn$ is critical for future work attempting to interface $Mn_3Sn$ with other quantum materials. Towards that end, we also grew (0001) $Mn_3Sn$ on the IC-technology compatible substrate Si(111) with a Ag(111) seed layer (Fig. S2F) and $(11\bar{2}0)$ $Mn_3Sn$ on $R-Al_2O_3$ substrate (Fig. S3).

It has been shown previously that hexagonal $Mn_3Sn$ is stable with excess Mn. Specifically, earlier work on bulk $Mn_{3+x}Sn_{1-x}$ found a maximum Mn excess of $x = 0.2$ could be achieved (*22, 23, 33*).

The extra Mn atoms replace Sn atoms and randomly occupy the Sn sites (i.e., $Mn_{3+x}Sn_{1-x}$, $x > 0$). Importantly, the replacement of non-magnetic Sn atoms with magnetic atoms is expected to alter real space spin ordering and impact the spacing of Weyl nodes in the system. DFT calculations also suggest that extra Mn may raise the Fermi energy $E_F$ (*22*). Therefore, doping with Mn should be an effective way to tune band structure topology and the Hall effects in $Mn_{3+x}Sn_{1-x}$.

As shown in the inset of Fig. 2A, for two distinct $x$ values, the (0002) XRD peaks are well separated, indicating distinct values of lattice constant $c$. The lattice constant $a$ is obtained from the ($20\bar{2}1$) peak at tilted geometry via the relation of $a = 4cd_{(20\bar{2}1)}/\sqrt{3(c^2 - d_{(20\bar{2}1)}^2)}$. As shown in Fig. 2D, lattice parameters $c$ and $a$ decrease as $x$ increases, supporting replacement of Sn atoms by smaller Mn atoms. Most importantly, we found that the amount of Mn excess can be greatly extended to $x \approx 0.55$ in our epitaxial films without the introduction of an impurity phase. This result sharply contrasts the small compositional window seen in bulk samples and is likely related to the non-equilibrium growth conditions that exist during vacuum deposition processes. As $x$ increases from 0.06 to 0.55, the $c$-axis lattice constant decreases from 4.57 Å to 4.53 Å. Similar dependence of lattice parameters as a function of $x$ is observed in (0001) $Mn_{3+x}Sn_{1-x}$ grown on Ag(111) seed layer on Si(111) wafer (Fig. 2D). The lattice parameters of ($11\bar{2}0$) $Mn_{3+x}Sn_{1-x}$ show different dependence on $x$, but the unit cell volume still decreases as $x$ increases (Fig. S3). The greatly extended compositional range for replacing Sn with magnetic Mn in thin films allows for the exploration of new and unusual physics in $Mn_3Sn$, as exemplified below.

**Kondo effect and gap opening by extended Mn doping**

Figure. 3A shows normalized resistance $\gamma(T)$ (= $R(T)/R(295K)$) as a function of temperature ($T$) for 60 nm (0001) $Mn_{3+x}Sn_{1-x}$ films. Strictly speaking, a rigorous identification of metallicity requires demonstration that resistance extrapolates to a finite value at zero temperature, which is possible even if $R$ is increasing as $T$ decreases throughout the measurement range (*34*). For simplicity, here we adopt the popular nomenclature that $\gamma(T)$ curves have a "metallic" ("insulating") characteristic if resistance decreases (increases) as $T$ decreases. For reference, we note that bulk $Mn_{3+x}Sn_{1-x}$ ($x < 0.15$) single crystals show metallic behaviors with carrier density $n$ about $2 \times 10^{22}$ cm$^{-3}$. The resistivity of such bulk crystals is about 250 μΩ·cm at room temperature along the $ab$-plane, indicating a mean-free-path on the order of 1 nm (*35*). Broadly speaking, the $\gamma(T)$ character of our films is metallic for compositions below $x \sim 0.40$ and insulating above. For $x = 0.06$, $\gamma$ decreases nearly linearly from room temperature to $T = 50$ K and reaches a residue value about 0.76 at $T = 5$ K. As $x$ increases to 0.11, metallic behavior remains, but with a larger $\gamma$ at 5 K, consistent with the elevated scattering from Mn dopants randomly replacing the centers of the hexagons in the Kagome plane. Note that the transport contributions from the buffer and capping layers are minimized by using highly resistive materials (Materials and Methods).

The strong correlation among Mn *3d* electrons (*22*) in $Mn_3Sn$ and its topological band structure suggest that $Mn_3Sn$ is an ideal platform to explore the interplay between Weyl and correlation physics. In this material, Mn moments in the Kagome lattice are strongly antiferromagnetically coupled to each other, leading to various spin orders such as a planar 120-degree spin orientation with negligibly small net magnetization (*25*). The substituting $Mn_{Sn}$ atom sitting at the hexagon center of the Kagome lattice could interact with its 6 neighbors via Heisenberg exchange. However, the vanishing of total magnetic moment of the 6 neighbors results in the same energy no matter what direction the center spin points toward, to lowest order. Therefore, substitutional Mn spins may behave as magnetic impurities exchange coupled to the *d*-like conduction band states on the Mn-Kagome sublattice, leading to Kondo physics if the correlation is considerable. Indeed, in the metallic regime, for example the $x = 0.27$ film, Fig. 3B shows a resistance upturn instead of a residue below the resistance minimum temperature $T_m = 25$ K. The -ln$T$ increase of resistance with decreasing temperature at $T < T_m$ and the deviation of ln$T$ scaling at lower temperature are characteristics of Kondo effect (*36*). For simplicity, we adopt the deviation point of ln$T$ scaling (Fig. 3B) as an estimation of the Kondo temperature $T_K$, which reflects the strength of Kondo interaction (*36*). The resistance increases as a result of Kondo scattering of magnetic impurities. As temperature further decreases, the Kondo scattering is inhibited due to the formation of a singlet state between a localized Mn moment and a conduction electron, which screens the magnetic moments of the impurities (*37*).

As *x* increases further and more Sn atoms are replaced by Mn atoms, $T_m$ and $T_K$ shift gradually to higher temperatures indicating a stronger Kondo effect. For $x = 0.39$, $T_m$ ($T_K$) is 47 K (23 K) (Fig. 3C). Above $x > 0.40$ (Fig. 3D, 3E), the resistance shows an upturn at much higher temperature (e.g., 192 K for $x = 0.44$, and $> 300$ K for $x = 0.50$), consistent with the transition to a Kondo insulator as Kondo effect develops coherence in the dense Kondo lattice (*14*). The ground state of such a system features a hybridization gap, on the order of a few meV to 10's of meV, near the Fermi energy (*14, 15*), and due to the featured band inversion therein, most Kondo insulators are topologically nontrivial (*16*). A prominent example is the topological Kondo insulator SmB$_6$, which shows a small insulator gap of about 3 meV and a saturation resistance plateau at low temperatures (*38*). The γ(*T*) curves of our highly doped films ($x > 0.40$) show a similar temperature dependence. As shown in Fig. 3E for $x = 0.55$, the resistance exhibits a -ln$T$ increase from room temperature to $T_K \sim 130$K, then increases at a slower rate and eventually reaches a near-saturated value at low temperature. Following the analysis in Ref. (*38*), we divide the measured conductance $G_{measured}$ into two parallel contributions: $G_{measured} = G_{T=5K} + G_{insulator}$, where $G_{T=5K}$ represents the near-saturated conductance at low temperature. $G_{T=5K}$ may originate from a topological surface state or an in-gap state (*39*). We plot $G_{insulator}$ as a function of $1/T$ from 300 K to 150 K (inset of Fig. 3E). The linearity indicates this analysis is appropriate and gives a gap value of 10.2 meV (*i.e.*, 118 K ≈ $T_K$), which corresponds well with reported gap values of other Kondo insulators (*15*). The gap opening at $x > 0.40$ is also evidenced by the decrease of carrier density as a function of temperature (Fig. S4) due to reduced thermal activation at low temperature.

We briefly note that the Kondo effect and gap opening are also observed in thicker (0001) Mn$_{3+x}$Sn$_{1-x}$ films, as well as (11$\bar{2}$0) oriented films, all with similar γ(*T*) curves (Fig. S4). Terahertz (THz) transmission measurements (Fig. 3F) at $T = 2$ K on a 60 nm (11$\bar{2}$0) Mn$_{3+x}$Sn$_{1-x}$ ($x = 0.47$) shows a prominent dip at 1.3 THz (~5.5 meV), which is less obvious in a metallic sample ($x = 0.13$). These THz transmission results echo the resistivity measurements and corroborate the notion of a gap opening. Even stronger evidence for the gap opening can be seen in THz Faraday rotation measurements discussed below.

**Resonance-enhanced THz Faraday rotation**

Here we present a striking consequence of the gap opening in the THz Hall effect (i.e., Faraday rotation). Fig. 4 shows the real ($\theta'_F$) and imaginary ($\theta''_F$) parts of the Faraday rotation, which are related to each other by Kromers-Kronig transform. The metallic sample ($x = 0.13$) shows a small

Faraday rotation on the order of a few mrad (Fig. 4A), which is similar to the observed value in a recent study on polycrystalline Mn$_3$Sn films (*40*). In sharp contrast, for the gapped sample ($x = 0.47$), the imaginary part of the Faraday rotation, which is related to the ellipticity ($\theta_F''$, Fig. 4B), shows a remarkable, resonance-enhanced peak at 1.33 THz (i.e., 5.6 meV, the aforementioned energy gap in the transmission measurements). According to the Kromers-Kronig relation, a resonance across the gap in the imaginary part corresponds to a sharp zero crossing (an inflection point) in the real part at 1.33 THz (Fig. 4A). We applied 6 T to align the magnetic domain first and then performed the Faraday rotation measurement by reducing the field from 4 T to 0 T. The resonance-enhanced $\theta_F'$ is 0.3 rad at 4 T, which is around 40 times larger than that in the recent study on metallic Mn$_3$Sn films (*40*). The gap induced resonance feature in optical rotations was also recently observed in a magnetic topological insulator (*41*). The resonance-enhanced THz Faraday rotation described here may represent an entirely new phenomenon in the world of the Kondo insulator, which was not observed in SmB$_6$ (*42*).

These above transport and terahertz results demonstrate a gap opening of Weyl semimetal by doping of magnetic Mn atoms whose 3*d* electrons have strong correlations, and thus a possible transition from Kondo effect to Kondo insulator in new class of topological matter. Interestingly, these effects are independent of the crystalline growth orientation.

**DC Hall resistances arising from Berry curvatures in reciprocal and real spaces**

One of the most salient transport features in bulk Mn$_3$Sn is the large spontaneous anomalous Hall resistance, arising from the Berry curvatures of the Weyl nodes (*22, 25*). Therefore, the Weyl nature of our Mn$_{3+x}$Sn$_{1-x}$ thin film can be checked by Hall measurements. The total Hall resistivity includes various contributions and can be described as $\rho_H = R_0 B + R_S \mu_0 M + \rho_H^{AF}$, where *M* is the magnetization, $R_0$ ($R_S$) is the ordinary (anomalous) Hall coefficient (both $R_0$ and $R_S$ are positive) and $\mu_0$ is the magnetic permeability. Unlike conventional ferromagnets such as Fe and Co where the Hall resistances only include the first two terms, there is additional term $\rho_H^{AF}$ which is almost independent of *B* or *M* (*25*). Since usual antiferromagnets do not exhibit measurable spontaneous Hall resistances, $\rho_H^{AF}$ presumably originates from the Berry curvatures in reciprocal or real space. The former contributes to an intrinsic anomalous Hall effect determined by the separation of Weyl nodes, while the latter manifests itself in topological Hall effect as seen in chiral magnets (*43*) and oxide heterostructures (*44*). It is important to note that exchange bias in Mn$_{3+x}$Sn$_{1-x}$/Py bilayers is observed at *T* = 5K (Fig. S5) in both metallic and gapped states. This is strong evidence that antiferromagnetism in our Mn$_{3+x}$Sn$_{1-x}$ films is robust over the whole compositional range we studied. Above the saturation field of the uncompensated spins (but below the spin-flop transition), $M = M_S^0 + \chi B/\mu_0$, where $\chi$ is the susceptibility and $M_S^0$ is the residual magnetization by extrapolating the high field *M* to *B* = 0. Therefore, above the saturation field, $\rho_H = (R_0 + \chi R_S)B + (R_S \mu_0 M_S^0 + \rho_H^{AF}) \equiv R_0^* B + \rho_{AHR}^*$, which includes a field-linear term and a field-independent term. In our films, independent of orientations, $M_S^0$ is on the order of 0.001 – 0.03 $\mu_B$/Mn at room temperature, close to the resolution limit of the magnetometers used in this work (Fig. S6 and Supplementary Note 2). In the following, we use bulk value $\chi \sim 3\times10^{-3}$ for our estimations (Supplementary Note 1).

We first consider the Hall effect in ($11\bar{2}0$) films for metallic states ($x < 0.4$). Figure S7A shows Hall resistance as a function of field for Mn$_{3+x}$Sn$_{1-x}$ ($x = 0.18$) at *T* = 300K. The Hall loops indicate $-\rho_H^{AF} \approx -\rho_{AHR}^* \gg R_S \mu_0 M_S^0$ (Fig. S7). The saturated Hall resistivity ($\rho_{AHR}^*$) at positive fields (e.g.,

B = 9 T) is negative. The negative AHR is a signature of the intrinsic contribution of Berry curvature in $k$-space as observed in bulk single crystal samples (*25*). Fig. 5A shows the $\rho^*_{AHR}$ ($\approx \rho^{AF}_H$) as a function of temperature for $(11\bar{2}0)$ films with different $x$. For $x = 0.10$ and $0.18$, $\rho^*_{AHR}$ is as high as -1.5 μΩ·cm at 300 K. As temperature decreases, $|\rho^*_{AHR}|$ decreases and a clear transition around 225 K is observed. The transition at $T_1 \approx 225$ K, which is also observed in bulk Mn$_3$Sn single crystals (*35*), indicates changes from triangular spin order at higher temperatures to spin glass or spin spiral at low temperatures. Between $T_N = 420$ K and $T_1$, a Weyl semimetal state exists due to the magnetic symmetry of triangular spin order (*22*). Below $T_1$, the intrinsic negative AHR decreases drastically due possibly to the disappearance of Weyl nodes in spin glass/spinal state. The $\rho_H$ ($T < T_1$) is only about -0.2 μΩ·cm and depends weakly on temperature. At $x = 0.26$ where Kondo physics manifests at low temperature (Figs. 3B and S4), a similar temperature dependence of $\rho^*_{AHR}$ is observed; $\rho^*_{AHR}$ remains negative at the whole temperature range despite slightly smaller values compared with those in films with $x = 0.10$ and $0.18$.

However, at $x > 0.40$ the gapped state is realized, $\rho^*_{AHR}$ shows drastically different temperature dependence. At $T = 300$ K, $\rho^*_{AHR}$ is only around -0.2 μΩ·cm, nearly an order-of-magnitude smaller than that of films with $x = 0.10$. This may be a direct consequence of a gap opening, which likely annihilates Weyl nodes and greatly reduces the Berry curvature. Alternatively, extended doping could also disrupt the triangular spin texture which protects the Weyl nodes. Interestingly, $\rho^*_{AHR}$ changes sign around $T_1$ (Fig. 5A). With extended doping ($x > 0.40$), the doped localized Mn moments start to form a lattice and disrupt the spin glass/spiral state at $T < T_1$. The doped Mn spins, along with Mn spins in the Kagome sites, are likely lying in the Kagome plane that is perpendicular to the film. These spins are 'polarized' by the out-of-plane field (in the Kagome plane) and give rise to positive conventional AHR $R_S\mu_0 M_S^0$ ($\leq \rho^*_{AHR}$, Supplementary Note 3). Note that a positive $\rho^{AF}_H$ may also contribute due to spin chirality if spins tilt towards the film plane.

$\rho^*_{AHR}$ in $(11\bar{2}0)$ Mn$_{3+x}$Sn$_{1-x}$ films are summarized in a *T-x* phase diagram shown in Fig. 5B. Large negative Hall resistance presumably induced by Weyl nodes exists at $T > 225$ K and $x < 0.40$. Significantly, driven by the Kondo effect with higher Mn doping, a gapped state (possibly a Kondo insulator) emerges at $x > 0.40$ and shows positive Hall resistance at $T < 225$ K. These results indicate magnetic doping is an effective way to tune electron correlations and Berry curvatures in the reciprocal space.

Although Kondo physics is similar in (0001) films, the Hall resistance shows different compositional dependence therein, due to the anisotropy of the Hall effect in these materials. In bulk single crystals ($x < 0.15$), the $M_S^0$ and $\rho^{AF}_H$ ($\rho^*_{AHR}$) is zero in the spin triangular state if the magnetic field is applied along [0001] direction (*25*). In the spin glass state, $\rho^*_{AHR}$ is reported to be positive (*25*). In (0001) Mn$_{3+x}$Sn$_{1-x}$ films, positive $\rho^*_{AHR}$ is observed from 300 K to 2 K. $\rho^*_{AHR}$ of the gapped state ($x > 0.40$, Fig. 5C and Fig. S8) is much smaller than that of the metallic state. $\rho^*_{AHR}$ of (0001) films in metallic state is significantly larger than the estimated upper-limit of $R_S\mu_0 M_S^0$, indicating that $\rho^{AF}_H$ is dominating (Supplementary Note 3). Furthermore, the major (9 T) and minor (1 T) Hall loops show distinct hysteresis (Fig. S4A), in sharp contrast to the loops in $(11\bar{2}0)$ films (Fig. S7A). The intrinsic $\rho^{AF}_H$ in (0001) films is likely field-induced topological Hall resistance ($\rho^{TH}$) (*45*) arising from the spin chirality in the real space (*27*). Considering a triangle spin structure shown in Fig. 5C, tilting of spins ($S_i$, i=1,2,3) towards *c*-axis gives a scalar spin chirality $S_1 \cdot (S_2 \times S_3)$ (*27*) which induces a spin Berry phase (magnetic flux) $\Phi = 2r\Phi_0$, where $\Phi_0$ is the flux quantum (2000 T·nm$^2$) and $r$ reflects the fraction of the flux quantum ($r$ is 1 for a magnetic skyrmion). The spin Berry phase gives rise a fictitious field $B_{eff} = \Phi/A = r \times 1.18 \times 10^5$ T, where $A$=0.034 nm$^2$ is the area of each triangle. Therefore, a tiny value of $r$ (Supplementary Note 3) can generate large $B_{eff}$ that induces a topological Hall resistivity $\rho^{TH} = R_0 B_{eff}$ (e.g., $\rho^{TH}$ =0.33 µΩ·cm for 60 nm films). In thicker films such as 120 nm with lower crystalline quality, tilting can easily take place resulting larger $r$. The larger $r$ thus corresponds to a larger $\rho^{TH}$ as echoed in the experiment (Fig. S8). In contrast with $\rho_H^{AF}$ in (11$\bar{2}$0) films, $\rho^{TH}$ increases from $T$ = 300 K to $T \approx T_1$ (225 K), but decreases noticeably at $T$ < 100 K, which suggests a spin glass transition at $T_g$ ~ 100 K (Supplementary Note 1). As mentioned above, for gapped states ($x$ > 0.40), near half of hexagon centers are occupied by magnetic Mn atoms with spins randomly orienting in the Kagome plane. These doped (random) spins contribute to the spin chirality ($\mathbf{S}_1 \cdot (\mathbf{S}_2 \times \mathbf{S}_3)$) with opposite signs so that the net spin chirality is (nearly) zero. These unusual Hall resistances in (11$\bar{2}$0) and (0001) films indicate that doping of magnetic Mn atoms is an effective knob to tune Berry curvatures in both real and reciprocal spaces.

**Unusual magnetoresistance**

Another important transport feature in Weyl semimetals is their negative magnetoresistance (NMR) due to the chiral anomaly. When a magnetic field is applied along the current direction, a chiral charge current is driven from one Weyl node to its counterpart with opposite chirality (*6*). This chiral current leads to additional electric conductivity and gives rise to NMR. The NMR effect is largest when $\mathbf{B} \parallel \mathbf{I}$ and vanishes when $\mathbf{B} \perp \mathbf{I}$ (*6, 22*). The NMR effect has been observed in bulk Mn$_3$Sn single crystals, where it is very small (~ 0.1% at $B$ = 9 T and $T$ = 300 K) (*22*). Moreover, the NMR decreases quickly as the energy difference ($\Delta E$) between Fermi energy and Weyl nodes increases. NMR is inversely proportional to $\Delta E^{\,2}$ and is reduced by ~60% as $\Delta E$ increases from 5 meV ($x$ = 0.06) to 8 meV ($x$ = 0.03) (*22*). As shown in Fig. 6A and Fig. S9A, a small ordinary positive magnetoresistance (PMR) due to Lorentz force is observed from $T$ = 300 K to $T$ = 2 K in our films for $x$ < 0.40. The absence of NMR is likely due to the larger $\Delta E$ in those films. Interestingly, for the gapped state (Figs. 6B and S9B), positive magnetoresistance (PMR) is observed at $T$ > 70K but NMR is observed at $T$ < 70 K inside the spin glass regime (i.e., $T_g$ ~ 70 K). The nearly linear NMR at low temperatures weakly depends on the orientation of the applied magnetic field. This isotropic NMR can originate from the suppressed Kondo scattering and/or spin dependent scattering as magnetic field polarizes the randomly oriented doped Mn spins at the center of the hexagon (Supplementary Note 4). Nevertheless, the NMR well correlate with the $\gamma(T)$ and Hall results induced by Mn doping, providing complementary transport consequences by the magnetic doping of Mn atoms.

**DISCUSSION**

In this work, we show that antiferromagnetic WSM Mn$_{3+x}$Sn$_{1-x}$ thin films with superior sample quality are a class of exciting materials to study the interplay between strong correlation, topology, and magnetism. By replacing Sn with magnetic Mn atoms, a Kondo effect, which often appears in strongly correlated metals, emerges, evolves, and leads to opening of a hybridization gap. As a consequence, Hall resistances inherent to Berry curvatures in the reciprocal space decreases. Strikingly, resonance-enhanced THz Faraday rotation emerges as the gap is opened. Our work calls for further studies of related materials. For example, one may increase electron localization (hence electron correlations) by doping atoms with 3*d* (e.g., Fe, Co, Cu) or 5*f*/4*f* (e.g., Gd) electrons (*14*). It is also interesting to tune the spin-orbit coupling by doping (especially isostructural and/or isoelectronic doping) of heavy elements (e.g., Pb).

One of the major challenges in antiferromagnetic spintronics is the electric detection of AFM spin order. In this respect, conventional collinear AFMs, which do not exhibit anomalous Hall resistances due to vanishingly small magnetization, are not good candidates for AFM spintronics (*28*). The rich non-collinear spin textures and the significant Hall resistances in the $Mn_3Sn$ family of compounds make it a promising contender for such applications. Moreover, the non-collinear spin orders and/or the topological aspects in $Mn_3Sn$ present new opportunities for spin-charge conversion and spin orbit torque switching of AFM spin orders. These $Mn_{3+x}Sn_{1-x}$ films with greatly extended compositional range, tunable strong correlations, unusual Hall resistances, and resonance-enhanced THz Faraday rotation, offer new prospects as exemplified in this work and may further propel the nascent field of topological AFM/THz spintronics and the development of novel spin-based devices.

## MATERIALS AND METHODS

The $Mn_{3+x}Sn_{1-x}$ films were synthesized by dc co-sputtering of Mn (99.99%) and Sn (99.99%) targets at a sputtering rate about 2 Å/s in a high vacuum magnetron sputtering system with base pressure better than $5 \times 10^{-8}$ torr. Each batch of films with compositional wedge (pure phase) was grown on a wafer with dimensions approximately 7mm×50mm. After growth, the wafer was cut into 13-14 pieces (each of them had size around 3.5mm×7mm) along the wedge direction for XRD measurements. Each piece was further cut for transport (~2mm×3.5mm, electrodes were made by wire bonder) and/or other characterizations. (0001) $Mn_{3+x}Sn_{1-x}$ films were grown at substrate temperature of $T_S \approx 300$ ºC on C-$Al_2O_3$ substrate with Pt seed layer and Si (111) substrate with Ag seed layer. (11$\bar{2}$0) $Mn_{3+x}Sn_{1-x}$ films were grown at substrate temperature of $T_S \approx 400$ ºC on R-$Al_2O_3$ substrate without seed layer. The samples presented in this work are listed in Table S1. The compositions of films were measured/calibrated on 150 nm (0001) $Mn_{3+x}Sn_{1-x}$ films by energy dispersive spectrometer (EDS) in a scanning electron microscope (SEM). For transport measurements on (0001) films, the buffer layer (Pt) is slightly doped (or the thickness is reduced) to increase the resistivity but retains its crystalline quality, so that the contribution from the buffer layer is minimized. The magnetization measurements were done in a Quantum Design PPMS system equipped with a vibrating sampling magnetometer (VSM) and a room temperature VSM. The DC transport measurements were done in a 9T cryogen-free Quantum Design PPMS system with transport rotator assembly. THz measurements were done in a home-built spectrometer that is coupled to a cryogen free 7 T split-coil superconducting magnet with a base temperature of 1.6 K.

## SUPPLEMENTARY MATERIALS

Section S1. Magnetic properties and spin orders in $Mn_3Sn$.

Section S2. Hall resistivity and carrier density analysis.

Section S3. Anomalous Hall resistivity in (11$\bar{2}$0) and (0001) films.

Section S4. Magnetoresistance in $Mn_{3+x}Sn_{1-x}$ films.

Table S1. List of samples.

**REFERENCES AND NOTES:**

1. K. v. Klitzing, G. Dorda, M. Pepper, New Method for High-Accuracy Determination of the Fine-Structure Constant Based on Quantized Hall Resistance. *Physical Review Letters* **45**, 494-497 (1980).
2. K. S. Novoselov, A. K. Geim, S. V. Morozov, D. Jiang, Y. Zhang, S. V. Dubonos, I. V. Grigorieva, A. A. Firsov, Electric Field Effect in Atomically Thin Carbon Films. *Science* **306**, 666 (2004).
3. C. L. Kane, E. J. Mele, Z2 Topological Order and the Quantum Spin Hall Effect. *Physical Review Letters* **95**, 146802 (2005).
4. A. A. Burkov, L. Balents, Weyl Semimetal in a Topological Insulator Multilayer. *Physical Review Letters* **107**, 127205 (2011).
5. X. Wan, A. M. Turner, A. Vishwanath, S. Y. Savrasov, Topological semimetal and Fermi-arc surface states in the electronic structure of pyrochlore iridates. *Physical Review B* **83**, 205101 (2011).
6. N. P. Armitage, E. J. Mele, A. Vishwanath, Weyl and Dirac semimetals in three-dimensional solids. *Reviews of Modern Physics* **90**, 015001 (2018).
7. A. A. Zyuzin, A. A. Burkov, Topological response in Weyl semimetals and the chiral anomaly. *Physical Review B* **86**, 115133 (2012).
8. K.-Y. Yang, Y.-M. Lu, Y. Ran, Quantum Hall effects in a Weyl semimetal: Possible application in pyrochlore iridates. *Physical Review B* **84**, 075129 (2011).
9. C. Y. Guo, F. Wu, Z. Z. Wu, M. Smidman, C. Cao, A. Bostwick, C. Jozwiak, E. Rotenberg, Y. Liu, F. Steglich, H. Q. Yuan, Evidence for Weyl fermions in a canonical heavy-fermion semimetal YbPtBi. *Nature Communications* **9**, 4622 (2018).
10. Y. Xu, J. Zhao, C. Yi, Q. Wang, Q. Yin, Y. Wang, X. Hu, L. Wang, E. Liu, G. Xu, L. Lu, A. A. Soluyanov, H. Lei, Y. Shi, J. Luo, Z.-G. Chen, Electronic correlations and flattened band in magnetic Weyl semimetal $Co_3Sn_2S_2$. *arXiv:1908.04561*, (2019).
11. V. Ivanov, X. Wan, S. Y. Savrasov, Topological Insulator-to-Weyl Semimetal Transition in Strongly Correlated Actinide System UNiSn. *Physical Review X* **9**, 041055 (2019).
12. Y. Shao, A. N. Rudenko, J. Hu, Z. Sun, Y. Zhu, S. Moon, A. J. Millis, S. Yuan, A. I. Lichtenstein, D. Smirnov, Z. Q. Mao, M. I. Katsnelson, D. N. Basov, Electronic correlations in nodal-line semimetals. *Nature Physics* **16**, 636–641 (2020).
13. J. Kondo, Resistance Minimum in Dilute Magnetic Alloys. *Progress of Theoretical Physics* **32**, 37-49 (1964).
14. P. Coleman, Heavy Fermions and the Kondo Lattice: a 21st Century Perspective. *arXiv:1509.05769*, (2015).
15. Z. Fisk, J. L. Sarrao, J. D. Thompson, D. Mandrus, M. F. Hundley, A. Migliori, B. Bucher, Z. Schlesinger, G. Aeppli, E. Bucher, J. F. DiTusa, C. S. Oglesby, H. R. Ott, P. C. Canfield, S. E. Brown, Kondo insulators. *Physica B: Condensed Matter* **206-207**, 798-803 (1995).
16. M. Dzero, J. Xia, V. Galitski, P. Coleman, Topological Kondo Insulators. *Annual Review of Condensed Matter Physics* **7**, 249-280 (2016).
17. D. Ma, H. Chen, H. Liu, X. C. Xie, Kondo effect with Weyl semimetal Fermi arcs. *Physical Review B* **97**, 045148 (2018).
18. A. K. Mitchell, L. Fritz, Kondo effect in three-dimensional Dirac and Weyl systems. *Physical Review B* **92**, 121109 (2015).
19. A. Principi, G. Vignale, E. Rossi, Kondo effect and non-Fermi-liquid behavior in Dirac and Weyl semimetals. *Physical Review B* **92**, 041107 (2015).

20. H.-H. Lai, S. E. Grefe, S. Paschen, Q. Si, Weyl-Kondo semimetal in heavy-fermion systems. *Proc Natl Acad Sci USA* **115**, 93-97 (2018).
21. S. Dzsaber, X. Yan, G. Eguchi, A. Prokofiev, T. Shiroka, P. Blaha, O. Rubel, S. E. Grefe, H.-H. Lai, Q. Si, S. Paschen, Giant spontaneous Hall effect in a nonmagnetic Weyl-Kondo semimetal. *arXiv:1811.02819*, (2018).
22. K. Kuroda, T. Tomita, M. T. Suzuki, C. Bareille, A. A. Nugroho, P. Goswami, M. Ochi, M. Ikhlas, M. Nakayama, S. Akebi, R. Noguchi, R. Ishii, N. Inami, K. Ono, H. Kumigashira, A. Varykhalov, T. Muro, T. Koretsune, R. Arita, S. Shin, T. Kondo, S. Nakatsuji, Evidence for magnetic Weyl fermions in a correlated metal. *Nat Mater* **16**, 1090-1095 (2017).
23. M. Ikhlas, T. Tomita, T. Koretsune, M.-T. Suzuki, D. Nishio-Hamane, R. Arita, Y. Otani, S. Nakatsuji, Large anomalous Nernst effect at room temperature in a chiral antiferromagnet. *Nat Phys* **13**, 1085-1090 (2017).
24. M. Kimata, H. Chen, K. Kondou, S. Sugimoto, P. K. Muduli, M. Ikhlas, Y. Omori, T. Tomita, A. H. MacDonald, S. Nakatsuji, Y. Otani, Magnetic and magnetic inverse spin Hall effects in a non-collinear antiferromagnet. *Nature* **565**, 627-630 (2019).
25. S. Nakatsuji, N. Kiyohara, T. Higo, Large anomalous Hall effect in a non-collinear antiferromagnet at room temperature. *Nature* **527**, 212-215 (2015).
26. N. H. Sung, F. Ronning, J. D. Thompson, E. D. Bauer, Magnetic phase dependence of the anomalous Hall effect in $Mn_3Sn$ single crystals. *Applied Physics Letters* **112**, 132406 (2018).
27. H. Ishizuka, N. Nagaosa, Spin chirality induced skew scattering and anomalous Hall effect in chiral magnets. *Science Advances* **4**, eaap9962 (2018).
28. T. Jungwirth, J. Sinova, A. Manchon, X. Marti, J. Wunderlich, C. Felser, The multiple directions of antiferromagnetic spintronics. *Nature Physics* **14**, 200-203 (2018).
29. L. Šmejkal, Y. Mokrousov, B. Yan, A. H. MacDonald, Topological antiferromagnetic spintronics. *Nature Physics* **14**, 242-251 (2018).
30. A. Markou, J. M. Taylor, A. Kalache, P. Werner, S. S. P. Parkin, C. Felser, Noncollinear antiferromagnetic $Mn_3Sn$ films. *Physical Review Materials* **2**, 051001 (2018).
31. Y. You, X. Chen, X. Zhou, Y. Gu, R. Zhang, F. Pan, C. Song, Anomalous Hall Effect–Like Behavior with In-Plane Magnetic Field in Noncollinear Antiferromagnetic $Mn_3Sn$ Films. *Advanced Electronic Materials* **5**, 1800818 (2019).
32. T. Higo, D. Qu, Y. Li, C. L. Chien, Y. Otani, S. Nakatsuji, Anomalous Hall effect in thin films of the Weyl antiferromagnet $Mn_3Sn$. *Applied Physics Letters* **113**, 202402 (2018).
33. P. J. Brown, V. Nunez, F. Tasset, J. B. Forsyth, P. Radhakrishna, Determination of the magnetic structure of $Mn_3Sn$ using generalized neutron polarization analysis. *Journal of Physics: Condensed Matter* **2**, 9409 (1990).
34. P. M. Vora, P. Gopu, M. Rosario-Canales, C. R. Pérez, Y. Gogotsi, J. J. Santiago-Avilés, J. M. Kikkawa, Correlating magnetotransport and diamagnetism of sp2-bonded carbon networks through the metal-insulator transition. *Physical Review B* **84**, 155114 (2011).
35. X. Li, L. Xu, L. Ding, J. Wang, M. Shen, X. Lu, Z. Zhu, K. Behnia, Anomalous Nernst and Righi-Leduc Effects in $Mn_3Sn$: Berry Curvature and Entropy Flow. *Physical Review Letters* **119**, 056601 (2017).
36. L. J. Zhu, S. H. Nie, P. Xiong, P. Schlottmann, J. H. Zhao, Orbital two-channel Kondo effect in epitaxial ferromagnetic L10-MnAl films. *Nature Communications* **7**, 10817 (2016).

37. H. Xue, Y. Hong, C. Li, J. Meng, Y. Li, K. Liu, M. Liu, W. Jiang, Z. Zhang, L. He, R. Dou, C. Xiong, J. Nie, Large negative magnetoresistance driven by enhanced weak localization and Kondo effect at the interface of $LaAlO_3$ and Fe-doped $SrTiO_3$. *Physical Review B* **98**, 085305 (2018).
38. Y. Li, Q. Ma, S. X. Huang, C. L. Chien, Thin films of topological Kondo insulator candidate $SmB_6$: Strong spin-orbit torque without exclusive surface conduction. *Science Advances* **4**, eaap8294 (2018).
39. Z. Xiang, Y. Kasahara, T. Asaba, B. Lawson, C. Tinsman, L. Chen, K. Sugimoto, S. Kawaguchi, Y. Sato, G. Li, S. Yao, Y. L. Chen, F. Iga, J. Singleton, Y. Matsuda, L. Li, Quantum oscillations of electrical resistivity in an insulator. *Science* **362**, 65 (2018).
40. T. Matsuda, N. Kanda, T. Higo, N. P. Armitage, S. Nakatsuji, R. Matsunaga, Room-temperature terahertz anomalous Hall effect in Weyl antiferromagnet $Mn_3Sn$ thin films. *Nature Communications* **11**, 909 (2020).
41. S. Patankar, J. P. Hinton, J. Griesmar, J. Orenstein, J. S. Dodge, X. Kou, L. Pan, K. L. Wang, A. J. Bestwick, E. J. Fox, D. Goldhaber-Gordon, J. Wang, S.-C. Zhang, Resonant magneto-optic Kerr effect in the magnetic topological insulator $Cr:(Sb_x,Bi_{1-x})_2Te_3$. *Physical Review B* **92**, 214440 (2015).
42. N. J. Laurita, C. M. Morris, S. M. Koohpayeh, P. F. S. Rosa, W. A. Phelan, Z. Fisk, T. M. McQueen, N. P. Armitage, Anomalous three-dimensional bulk ac conduction within the Kondo gap of $SmB_6$ single crystals. *Physical Review B* **94**, 165154 (2016).
43. S. X. Huang, C. L. Chien, Extended Skyrmion Phase in Epitaxial FeGe(111) Thin Films. *Physical Review Letters* **108**, 267201 (2012).
44. W. Wang, M. W. Daniels, Z. Liao, Y. Zhao, J. Wang, G. Koster, G. Rijnders, C.-Z. Chang, D. Xiao, W. Wu, Spin chirality fluctuation in two-dimensional ferromagnets with perpendicular magnetic anisotropy. *Nature Materials* **18**, 1054–1059 (2019).
45. P. K. Rout, P. V. P. Madduri, S. K. Manna, A. K. Nayak, Field-induced topological Hall effect in the noncoplanar triangular antiferromagnetic geometry of $Mn_3Sn$. *Physical Review B* **99**, 094430 (2019).

**Acknowledgements: Funding:** X.H. and L.W. acknowledge support by the Amy Research Office under the Grant W911NF1910342. The Penn NSF MRSEC DMR-1720530 supported part of the magneto-terahertz instrumentation development (L.W.) and the electrical transport measurements (S.H.P. and J.M.K). J.Z. acknowledges the finical support by the U.S. Department of Energy (DOE), Office of Science, Basic Energy Sciences (BES) under Award No. DE-SC0020221. J.W. acknowledges the support by the Department of Energy, Office of Science, Basic Energy Sciences, Materials Sciences and Engineering Division, under contract DE-AC02-76SF00515. R.F.N. acknowledges support from the National Research Council Research Associateship Program. Research performed in part at the NIST Center for Nanoscale Science and Technology. P.K. and W.W. acknowledges support by the National Science Foundation under the Grant DMR-1905783.
**Author contributions:** S.X.H conceived the project. D.K. synthesized the films and performed the XRD measurements. T.R.T, S.H.P, and J.M.K. performed the DC transport measurements. X.H. and L.W performed the THz measurements and analyzed the data. J.W., P.K., and W.W. performed magnetic measurements. R.F.N. did HR-XRD and EDS measurements. S.X.H. analyzed the data with assistance from T.R.T and D.K., and inputs from all authors. S.X.H. and J.Z. discussed results and gave explanations. S.X.H wrote the manuscript with revisions from J.Z., J.M.K., L.W., and R.F.N. and comments from all authors. **Competing interests:** The authors declare that they have no competing interests. **Data and materials availability:** All data needed to evaluate the conclusions in the paper are present in the paper and/or the Supplementary Materials. Additional data related to this paper may be requested from the authors.

**FIGURES**

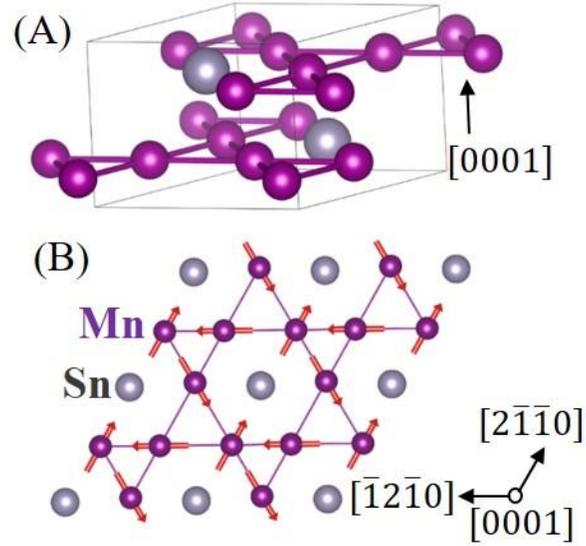

**Fig. 1. Crystal structure and spin structure of Mn$_3$Sn.** (A) Crystal structure of Mn$_3$Sn which consists of stacked kagome Mn$_3$Sn layers, (B) triangular spin structure in the kagome layer (*ab*-plane).

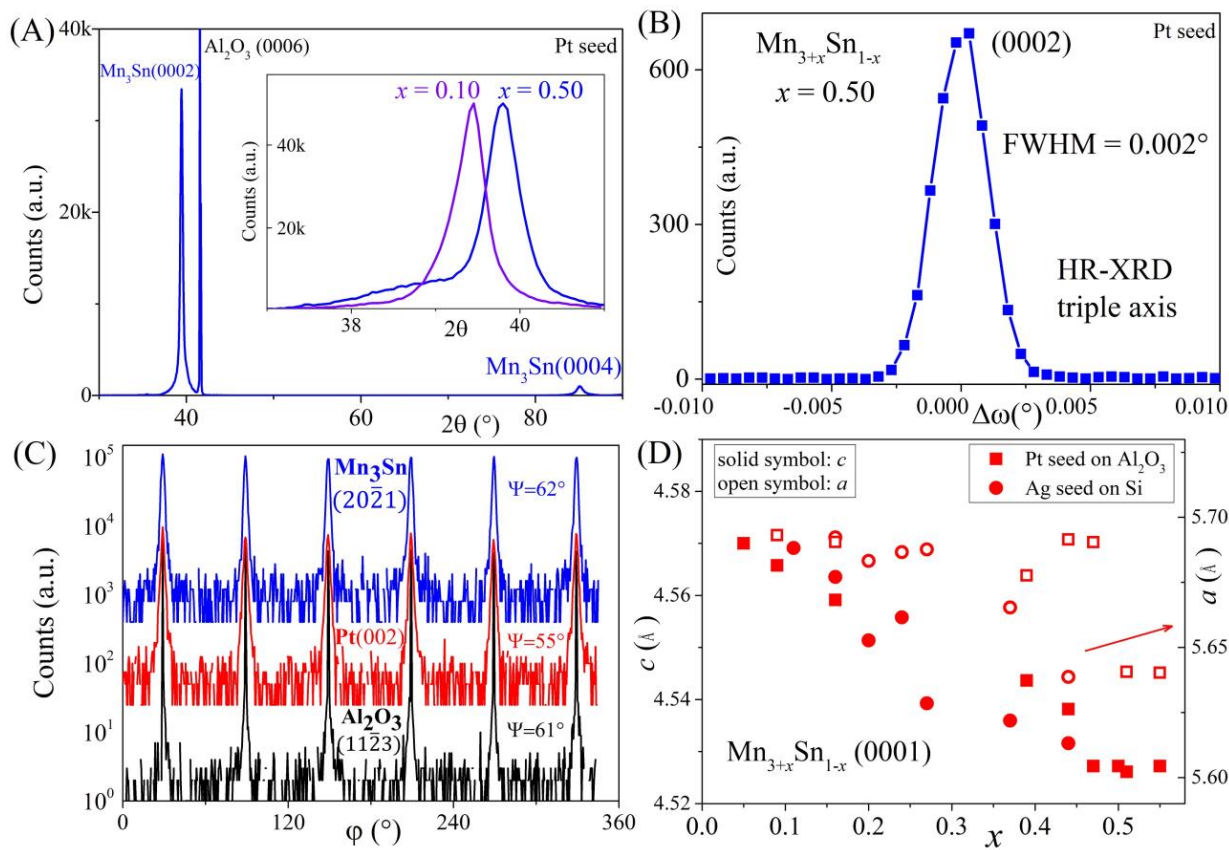

**Fig. 2. Structure characterization and extended compositional range of (0001) films.** (A) Representative XRD 2θ scan of (0001) Mn₃Sn films. inset: XRD 2θ scan of (0001) films with different compositions. (B) Triple axis HR-XRD ω scan (rocking curve) of Mn$_{3+x}$Sn$_{1-x}$ (0002) peak ($x$=0.50). (C) In-plane φ scans of Mn₃Sn (20$\bar{2}$1), Pt (002), and Al₂O₃ (11$\bar{2}$3) planes with tilt angles Ψ showing in the figure. (D) Lattice constants $c$ (solid symbols) and $a$ (open symbols) as a function of compositions.

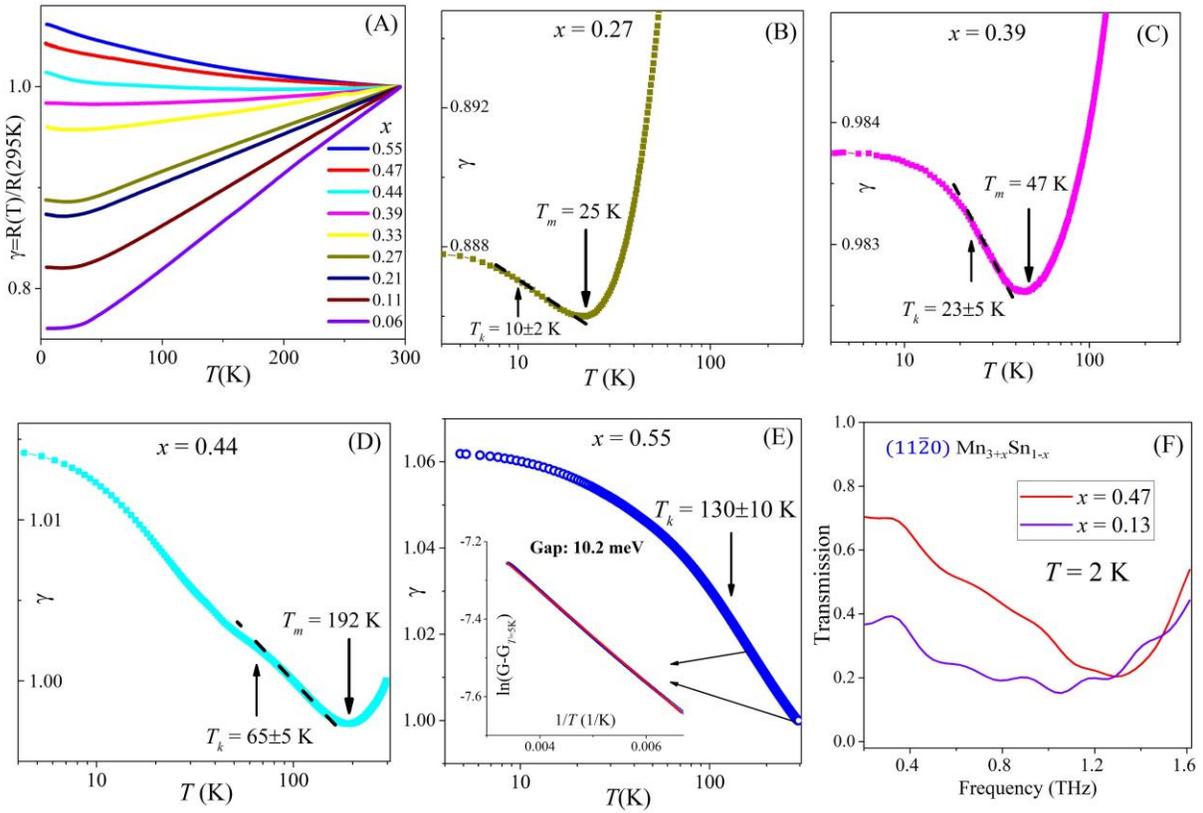

**Fig. 3. Evolution of Kondo effect and gap opening in $Mn_{3+x}Sn_{1-x}$ films.** Normalized resistance γ as a function of temperature for various $x$ (A), for (B) $x = 0.27$, (C) $x = 0.39$, (D) $x = 0.44$, and (E) $x = 0.55$, respectively. Inset of (E): $Ln(G-G_{T=5K})$ as a function of $1/T$, linear fit (red line) gives a gap value of 10.2 meV. (F) Transmission of $x = 0.47$ (red) and $x = 0.13$ (violet) samples as a function of frequency.

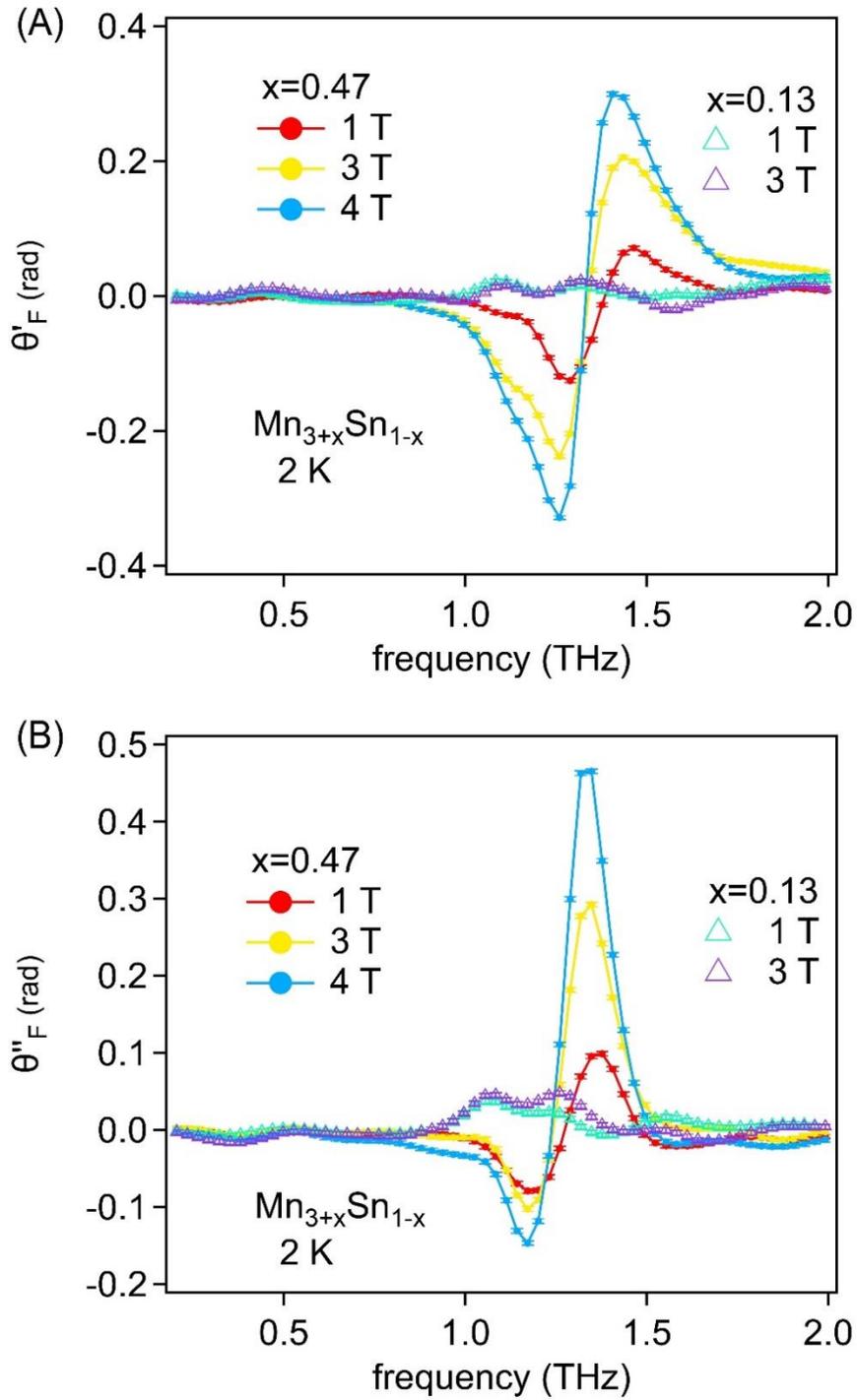

**Fig. 4. Resonance-enhanced THz Faraday rotation in ($11\bar{2}0$) Mn$_{3+x}$Sn$_{1-x}$ films.** Real part (A) and imaginary part (B) of the Faraday rotation as a function of frequency at different fields for $x = 0.47$ and $x = 0.13$ at $T = 2$ K.

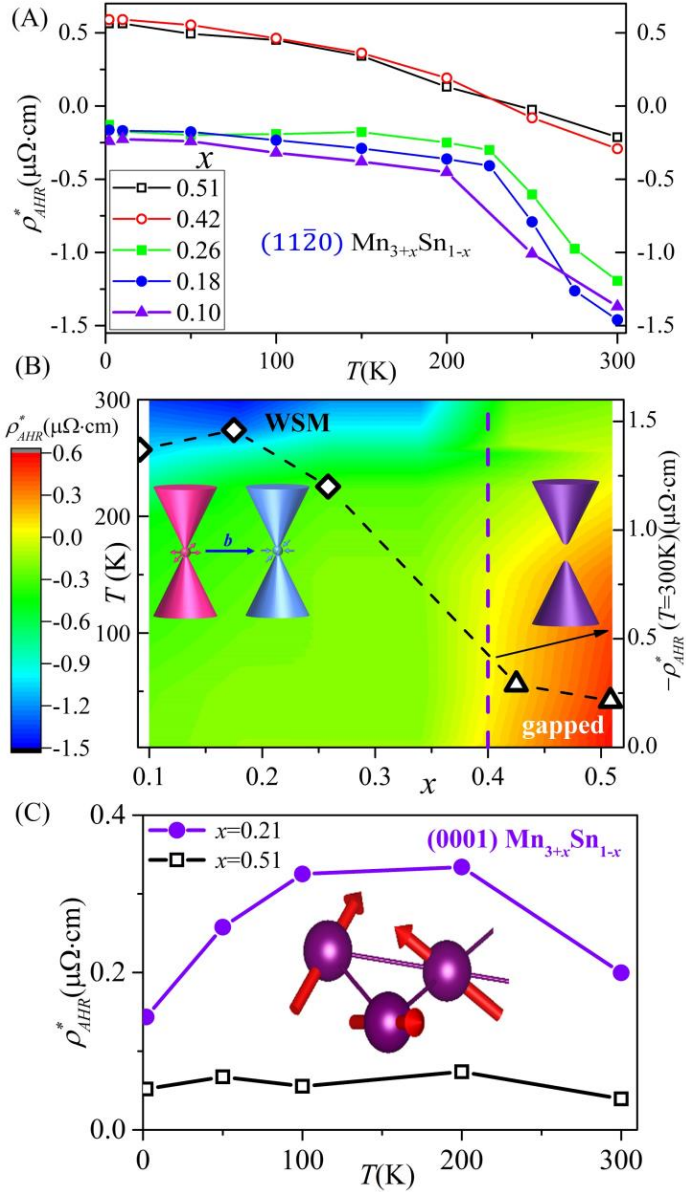

**Fig. 5. Anomalous Hall effects and phase diagram of Mn$_{3+x}$Sn$_{1-x}$ films.** (A) Anomalous Hall resistivity $\rho^*_{AHR}$ as a function of temperature for different compositions for $(11\bar{2}0)$ films. (B) Colored contour map of $\rho^*_{AHR}$ in $T$-$x$ plane for $(11\bar{2}0)$ films. Right $y$ axis: - $\rho^*_{AHR}$ ($T$=300K) as a function of $x$. Inset of (B): schematic diagrams of Weyl cones with opposite chirality and gapped cone. (C) Anomalous Hall resistivity $\rho^*_{AHR}$ of (0001) films as a function of temperature for $x$ = 0.21 (solid circles) and $x$ = 0.51 (open squares), respectively.

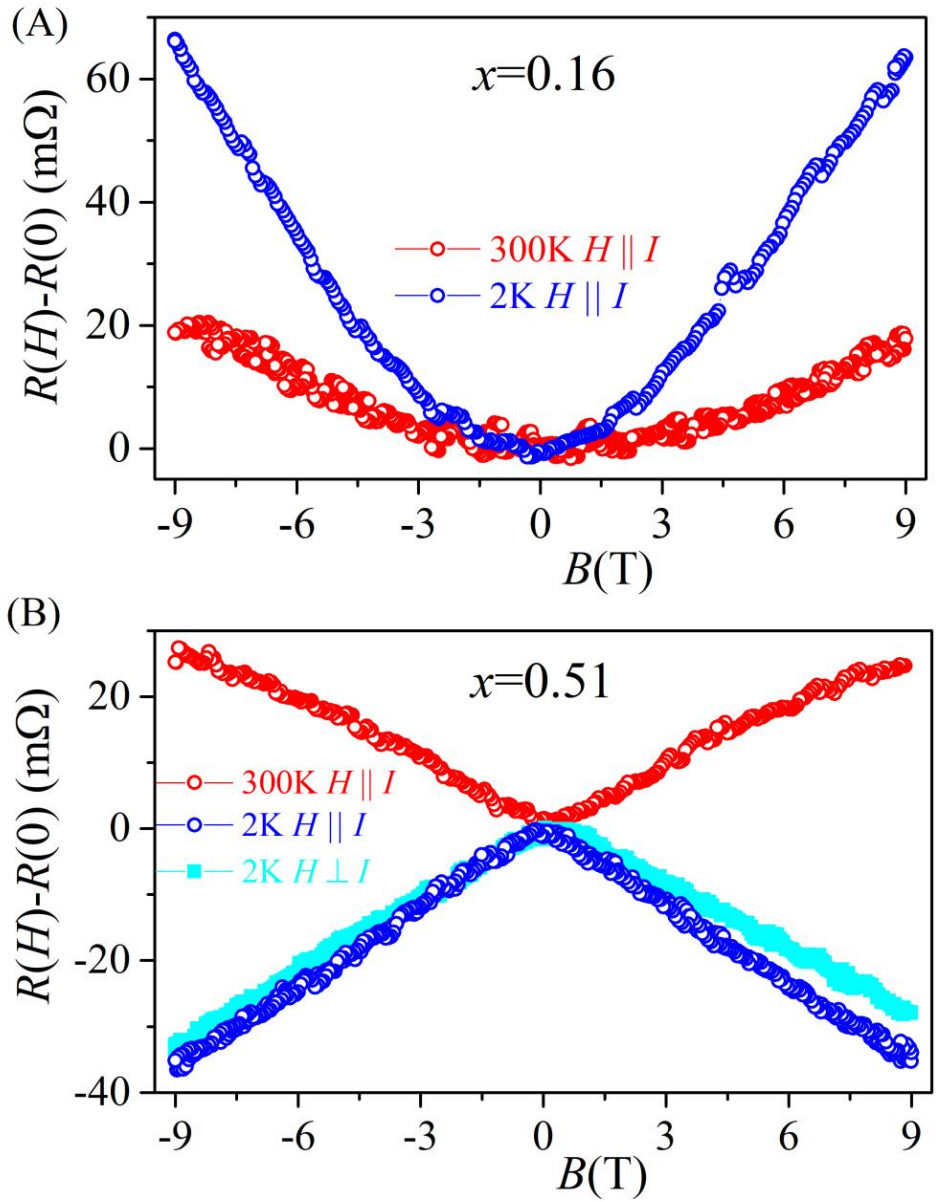

**Fig. 6. Magnetoresistances of (0001) Mn$_{3+x}$Sn$_{1-x}$ films.** Resistance change [$R(H)-R(H=0)$] as a function of field for (A) $x = 0.16$, and (B) $x = 0.51$, at $T = 2K$ (blue) and $T = 300$ K (red).